# Highly efficient star formation fueled by cold stream accretion in NGC 5253

Turner, J. L.[1], Beck, S. C.[2], Benford, D. J.[3], Consiglio, S. M.[1], Ho, P. T. P.[4], Kovács, A.[5], Meier, D. S.[6], & Zhao, J.-H.[7]

1. Department of Physics and Astronomy, UCLA, Los Angeles CA 90095-1547 USA
2. Department of Physics and Astronomy, University of Tel Aviv, Ramat Aviv, Israel
3. Observational Cosmology Lab., Code 665, NASA at Goddard Space Flight Center, Greenbelt, MD 20771, USA
4. Academia Sinica Astronomy and Astrophysics, Taipei, Taiwan
5. Department of Physics, Caltech, Pasadena, CA 91125, USA; Institute for Astrophysics, University of Minnesota, Minneapolis, MN 55405, USA
6. Department of Physics, New Mexico Institute of Mining and Technology, Socorro, NM 85723 USA; National Radio Astronomy Observatory
7. Harvard-Smithsonian Center for Astrophysics, 60 Garden Street, Cambridge, MA 02138 USA

**Gas clouds in present-day galaxies are inefficient at forming stars. Low star formation efficiency is a critical parameter in galaxy evolution: it is why stars are still forming nearly fourteen billion years after the Big Bang [1] and why star clusters generally do not survive their births, instead dispersing to form galactic disks, halos, or bulges [2]. Yet the existence of ancient massive star clusters in the Milky Way, globular clusters, suggests that efficiencies were higher when they formed ten billion years ago. A local dwarf galaxy, NGC 5253, has a young star cluster that may provide an example of highly efficient star formation [3]. Here we report the detection of the J= 3-2 rotational transition of CO at the location of the massive cluster. The gas cloud is hot, dense, quiescent, and extremely dusty. Its gas-to-dust ratio is lower than the Galactic value, which we attribute to dust enrichment by the embedded star cluster. Its star formation efficiency exceeds 50%, ten times higher than clouds in the Milky Way: this cloud is a factory of stars and soot. We suggest that high efficiency results from the force-feeding of star formation by a streamer of gas falling into the galaxy.**

The Submillimeter Array image of NGC 5253, shown in Figure 1, reveals a bright CO(3-2) source coincident with the giant cluster and its "supernebula" [4]. "Cloud D" [3] is one of only two molecular clouds detected within the galaxy; the second cloud is smaller and located ~5" (90 pc) to the southwest. A "streamer" of gas extending along the minor axis is also detected in CO(3-2). This streamer, previously detected in lower J CO lines, appears to be falling into the galaxy near the supernebula [3,5]. Both the streamer and Cloud D emit 870μm continuum emission, as shown in Figure 2. Also shown is an image of 350μm continuum, in which both Cloud D and the streamer are detected.

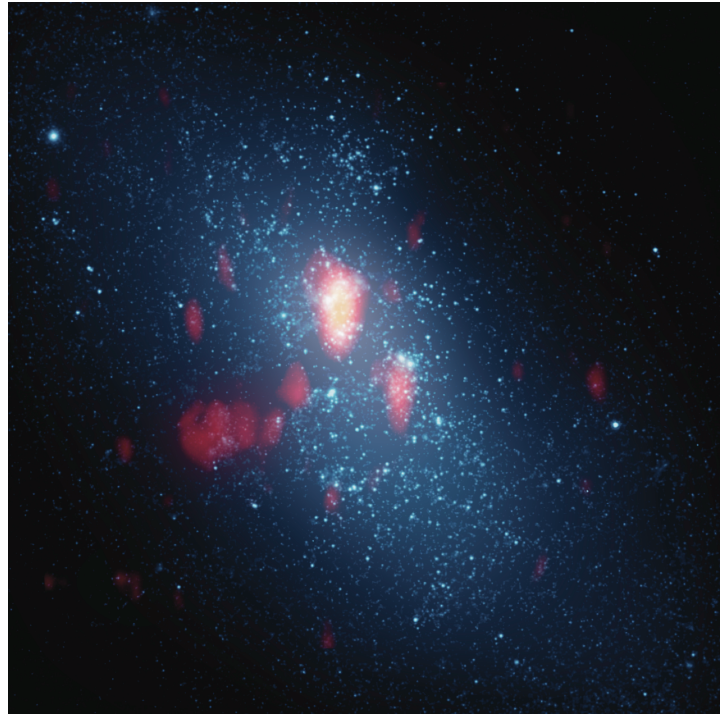

**Figure 1. CO J=3-2 emission in NGC 5253.** The SMA CO(3-2) integrated line intensity, in red, is shown atop a λ814nm Hubble Space Telescope image. The SMA beam is 4" x 2" (74 pc x 37 pc). The field covers 40" x 40" (740 pc x740 pc), north up, east left. Image registration is to < 1". The CO streamer coincides with the optical dust lane to the east. The massive star cluster is located at the bright CO peak, Cloud D; it is embedded [8,9], and not visible here. Cloud F is to the southwest of Cloud D.

The molecular gas in Cloud D is hot. This is clear from the increase in brightness from CO(2-1) [3] to CO(3-2). The intensity ratio of the two lines is $I_{32}/I_{21} = 2.6\pm0.5$ ($I_{line} = \int T_{line}$ dv). This ratio is nonthermal, although the thermal limit of 2.25 (~$v^2$) is within the uncertainties, and is what we adopt. Non-LTE modeling of this ratio using RADEX [6] indicates a minimum kinetic temperature of $T_K > 200$ K for the 1σ lower limit, and $T_K > 350$ K for the adopted value of $I_{32}/I_{21} = 2.25$ (see Methods). The high gas temperature is consistent with a thermal origin for $H_2$ 2.2μm emission in the region [7]. Cloud D appears to be a Photon-Dominated Region (PDR), heated by ultraviolet radiation from the several thousand cluster O stars in the cluster [4,8]. The CO(3-2)-emitting gas is dense, with $n_{H2}$ ~4.5±0.5 x$10^4$ cm$^{-3}$.

By contrast with Cloud D, the streamer consists of more typical cool giant molecular clouds. Its value of $I_{32}/I_{21} = 1.0 \pm 0.3$ is consistent with optically thick emission, for which

RADEX models allow temperatures as low as $T_K \sim 15\text{-}20$ K, and number densities $n_{H2} \sim 3.5\text{-}4 \times 10^4$ cm$^{-3}$. The mass of the streamer is $M_{H2} \sim 2 \times 10^6$ M$_\odot$[3], which is 1-2% of the stellar mass of the galaxy. The streamer is molecular, dense, and primed for star formation, even before entering the galaxy. This is unlikely to be a primordial collapsing filament [9], but previously enriched gas.

The star formation efficiency of a cloud or region can be defined as $\eta = M_{stars}/(M_{gas}+M_{stars})$, where $M_{stars}$ is the stellar mass and $M_{gas}$ the molecular gas mass. $M_{gas}$ can be hard to define for star-forming regions within giant molecular clouds, but the association of the isolated Cloud D with the supernebula gives us an opportunity to calculate $\eta$ directly for the giant molecular cloud giving birth to this massive star cluster—if we can determine the mass of Cloud D.

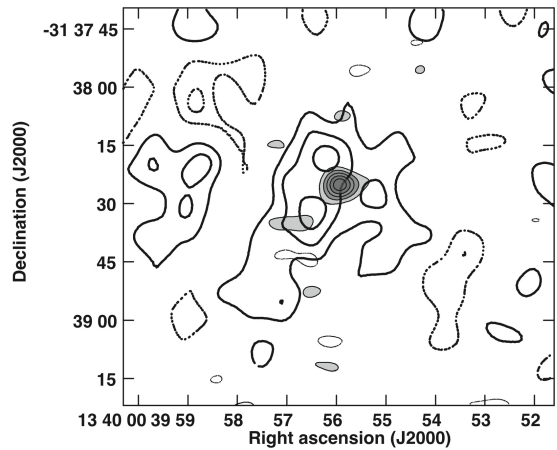

**Figure 2. Dust and gas in NGC 5253.** SMA image of continuum dust emission at 870μm (greyscale), with 350μm dust continuum emission from SHARC at the Caltech Submillimeter Observatory (contours) superimposed. The SMA continuum image has been smoothed to 6" resolution to show emission from the streamer. The SHARC image has been smoothed to 12.7"; contours are 2σ. Coordinates of the SHARC image are uncertain to ~5" (see Methods).

CO is often used to estimate molecular gas mass, but is unreliable in NGC 5253. Here we use the width of the CO line to determine a gas mass for Cloud D based on dynamical considerations. The CO linewidth is $\sigma = 9.2 \pm 0.6$ km/s, based on a Gaussian fit. The cloud dimensions, deconvolved from the beam, are 2.8" x 1.5"±0.1" (52 pc x 28 pc). The virial mass is $M_{vir} < 1.8^{+0.2}_{-0.7} \times 10^6$ M$_\odot$ for Cloud D, with uncertainties due to the unknown internal mass distribution (see Methods). We assume that the linewidth is gravitational, with no winds or outflow, hence this is an upper limit to $M_{vir}$. The virial mass includes both gas and stars, but we can constrain the stellar mass. The mass in stars exciting the supernebula, $M_{stars}$, can be predicted from the Lyman continuum rate of $N_{Lyc} = 7 \pm 2 \times 10^{52}$ s$^{-1}$ [6] and Brackett γ equivalent width of 255Å.[10] The star cluster has mass $M_{stars} = 1.1^{+0.7}_{-0.2} \times 10^6$ M$_\odot$ (see Methods). We then obtain a gas mass $M_{gas} = M_{vir} - M_{stars} = 7 \pm 4 \times 10^5$ M$_\odot$, if the cluster is embedded in the cloud. This treatment assumes that the CO kinematics trace all of the cloud and that there is no extensive layer of H$_2$ without CO; however, the dust continuum and CO sizes are nearly identical, consistent with the dust and gas mass being contained within Cloud D.

Other methods to estimate $H_2$ mass are problematic for Cloud D. CO(3–2) is optically thin, but the excitation temperature is not determined, nor is CO/$H_2$ known. The line strength of 41 ± 8 Jy km/s gives $M_{CO}$= 4 ± 1 $M_\odot$ (T / 200 K). For a Galactic abundance ratio of [CO]/[$H_2$] =8.5 x $10^{-5}$, the $H_2$ mass would be $M_{H2}$ = 5 x $10^4$ $M_\odot$ (T / 200 K). This is a factor of ten less than $M_{gas}$ derived above. Intense radiation fields and high temperature will affect the chemistry and the CO relative abundance.

Dust continuum emission can trace gas mass, but it is also unreliable in Cloud D. The 870 μm continuum flux density for Cloud D is $S_{870\mu m}$ = 72 ± 10 mJy, consistent with previous measurements [11], of which free-free emission [8] contributes $S_{870\mu m}^{ff}$= 38 ± 4 mJy, leaving dust emission $S_{870\mu m}^{dust}$ = 34±14 mJy. Our 350 μm image gives $S_{350\mu m}^{dust}$ =1.0 ± 0.2 Jy for Cloud D, consistent with the 870μm flux. Adopting the dust opacity of the Large Magellanic Cloud [12] (see Methods), we find an observed dust mass, $M_{dust}$ = 1.5 ± 0.2 x $10^4$ $M_\odot$ (T / 45 K)$^{-1}$. To obtain gas mass, we need a gas-to-dust ratio (GTD). Scaled as the oxygen abundance of NGC 5253, 0.2-0.3 solar [13,14], GTD~650, within the range 340-1200 inferred for the Magellanic clouds [15,16,17]. The observed dust mass is five times larger than the ~3000 $M_\odot$ of dust expected for a 2 x $10^6$ $M_\odot$ gas cloud at GTD = 650. If instead we compare the observed dust mass to the dynamical estimate of gas mass, we derive GTD ~47 for the embedded cluster, and in the unlikely case that the cluster is not embedded, ~120. Dust is an expected result of mass loss from massive, short-lived stars. Stellar models indicate that a cluster of massive stars of age 4.4 Myr, consistent with recombination line equivalent widths, will expel 20,000-30,000 $M_\odot$ of elements carbon, oxygen, silicon, magnesium, and iron depending on the cluster mass and initial mass function, of which ~30-50% will be in the form of dust (see Methods). To produce the amounts of dust and ionizing photons observed given the upper limit imposed by the dynamical mass suggests that the stellar initial mass function is top-heavy, with lower mass cutoff ≥ 2-3 $M_\odot$ (see Methods.) The star cluster has probably produced most of the dust. To infer a gas mass based on the observed dust emission for Cloud D from a GTD scaled to the global metallicity of NGC 5253 without accounting for in situ dust production, as has been done for other galaxies[18], would give an erroneously high gas mass and underestimate the star formation efficiency.

Given the peculiarities of Cloud D, the most reliable gas mass is dynamical. We use this mass to calculate star formation efficiency, η. A lower limit to η obtains if the gravitational mass is all gas, no stars, so that η = $1.1^{+.7}_{-.2}$ x $10^6$ $M_\odot$ / ($1.1^{+.7}_{-.2}$ + $1.8^{+0.2}_{-0.7}$) x $10^6$ $M_\odot$ ~$38^{+24}_{-7}$%. Even in this case η significantly exceeds the η <1% of Galactic giant molecular clouds and the highest efficiencies of η ~15-20% seen in individual cloud cores[19] in the Galaxy. However, it is virtually certain that the star cluster is located within the cloud, given the subarcsecond positional coincidence of nebular emission and CO, the precise kinematic coincidence of nebular H53α [8] and CO line centroids, and the high extinction to the cluster. Thus the stellar mass also contributes to the linewidth. For the more realistic case that the cluster is embedded within Cloud D, η = ($1.1^{+.7}_{-.2}$) x $10^6$ $M_\odot$ / $1.8^{+0.2}_{-0.7}$ x $10^6$ $M_\odot$ = $61^{+84}_{-16}$%. This value exceeds even the canonical ~50% (see Methods) needed to allow a star cluster to survive in its current bound state with rapid gas dispersal. If dust competes with gas for ultraviolet photons, the Lyman continuum rate and stellar mass have been underestimated, and η is even higher. If there are winds or outflows contributing to the CO linewidth, η is higher. The large dust mass favors larger mass cluster models for which η is higher. These values for η are uncertain, but free of the systematics due to standard assumptions such as gas-to-dust ratio, relative CO abundance, or CO conversion factor. The star formation efficiency of Cloud D is unusually high, implying gas consumption timescales of ~10 Myr.

A measurement of star formation efficiency is a snapshot in time; η could be high because the gas has been incorporated into stars or because the young stars have already dispersed the gas. How and when a young star cluster disperses its gas is key to its survival [2,20]. In the case of Cloud D, the youth of the embedded star cluster [10], its positional coincidence with the cloud, lack of evidence for supernovae [21], and small CO(3-2) linewidth strongly constrain gas dispersal models. Apparently not much has yet escaped this cloud.

Cloud D is a strange molecular cloud: hot, dusty, and small in mass relative to its young star cluster. It is found in a dark-matter-dominated galaxy. Its unusual properties may indicate a mode of star formation different from that observed in disk galaxies, including luminous infrared galaxies. Models of stochastic star formation for turbulently-supported giant molecular clouds in our Galaxy suggest that star formation efficiencies are 1% in a free-fall time [22], which implies that the ultimate efficiency can be limited if star formation is quenched by massive stellar feedback. An extended period of star formation might be facilitated if Cloud D is compressed by an external influence, as for example by a streamer of gas force-fed into the star-forming region by the galactic potential. Our data in NGC 5253 could support such a model. The streamer contains ~2 x $10^6$ $M_\odot$ of gas extending ~200-300 pc along the minor axis, entering the galaxy at a rate of ~20 pc/Myr. The streamer can fuel star formation at the present rate of ~0.1-0.2 $M_\odot$/yr for the next 10 million years. This dwarf spheroidal galaxy is not rotationally supported[23]; multiple accreting streams from its extensive HI halo[24,25] could be responsible for its global dynamics and morphology[26] as well as its spheroidal system of massive star clusters spanning Gyr in age [27,28,29]. NGC 5253 may illustrate a new mode of highly efficient star cluster formation triggered by cold stream accretion[30].

# References

1. Kennicutt, R. C., Jr. The Global Schmidt Law in Star-forming Galaxies. *Astrophys. J*, **498**, 541-552 (1998).

2. Bastian, N. & Goodwin, S. P. Evidence for the strong effect of gas removal on the internal structure of young stellar clusters. *Mon. Not. R. Astron. Soc.,* **369**, L9-L13 (2006).

3. Meier, D. S., Turner, J. L., & Beck, S. C. Molecular gas and the young starburst in NGC 5253 Revisited. *Astron. J.,* **124**, 877-885 (2002).

4. Turner, J. L., Beck, S. C., & Ho, P. T. P. The radio supernebula in NGC 5253. *Astrophys. J*, **532**, L109-L112 (2000).

5. Turner, J. L., Beck, S. C., & Hurt, R. L. A CO Map of the Dwarf Starburst Galaxy NGC 5253. *Astrophys. J.,* **474**, L11-L14 (1997).

6. van der Tak, F. F. S., Black, J. H., Schöier, F. L., Jansen, D. J., van Dishoeck, E. F. A computer program for fast non-LTE analysis of interstellar line spectra. With diagnostic plots to interpret the line intensity ratios. *Astron. & Astrophys.,* **468**, 627-635 (2007).

7. Cresci, G., Vanzi, L., Sauvage, M., Santagelo, G., & van der Werf, P. Integral-field near-infrared spectroscopy of two blue dwarf galaxies: NGC 5253 and He 2-10. *Astron. & Astrophys.,* **520**, A82 (2010).

8. Rodríguez-Rico, C. A., Goss, W. M., Turner, J. L., & Gómez, Y. VLA H53α observations of the central region of the super star cluster galaxy NGC 5253, *Astrophys. J.*, **670**, 295-300 (2007).

9. Gray, W. J., & Scannapieco, E. Thermal and chemical evolution of collapsing filaments. *Astrophys. J.*, **768**, 174 (2013).

10. Alonso-Herrero, A., et al. Obscured star formation in the central region of the dwarf galaxy NGC 5253. *Astrophys. J.,* **612**, 222-237 (2004).

11. Hirashita, H. Properties of free-free, dust and CO emissions in the starbursts of blue compact dwarf galaxies. *Mon. Not. R. Astron. Soc.*, **429**, 3390-3401 (2013).

12. Galliano, F. et al. Non-standard grain properties, dark gas reservoir, and extended submillimeter excess, probed by Herschel in the Large Magellanic Cloud. *Astron. Astrophys.,* **536**, A88 (2011).

13. Kobulnicky, H. A. , Skillman, E.D., Roy, J-R., Walsh, J.R., & Rosa, M.R., HST FOS Spectroscopy of localized chemical enrichment from massive stars in NGC 5253, *Astrophys. J.,* **477**, 679-692 (1997)

14. López-Sánchez, Á. R., Esteban, C., García-Rojas, J., Peimbert, M., & Rodríguez, M. The localized chemical pollution in NGC 5235 revisited: results from deep echelle spectrophotometry. *Astrophys. J.*, **656**, 168-185 (2007).

15. Bot, C., Boulanger, F., Lagache, G., Cambrésy, L., & Egret, D. Multi-wavelength analysis of the dust emission in the Small Magellanic Cloud, *Astron. Astrophys.*, **423**, 567-577 (2004).

16. Gordon, K. et al. The dust-to-gas ratio in the Small Magellanic Cloud Tail, *Astrophys. J.,* **690**, L76-L80 (2009).

17. Roman-Duval, J. et al. Dust/gas correlations from Herschel observations. *Astron. Astrophys,* **518**, L74 (2010).

18. Shi, Y., et al. Inefficient star formation in extremely metal poor galaxies. *Nature*, **514**, 335-338 (2014).

19. Olmi, L., & Testi, L. Constraints on star formation theories from the Serpens molecular cloud and protocluster. *Astron. & Astrophys.,* **392**, 1053-1068 (2002).

20. Smith, R., Goodwin, S., Fellhauer, M., & Assmann, P. Infant mortality in the hierarchical merging scenario: dependence on gas expulsion time-scales. *Mon. Not. R. Astron. Soc.,* **428**, 1303-1311 (2013).

21. Beck. S. C., Turner, J. L., Ho, P. T. P., Lacy, J. H., & Kelly, D.M. The central star cluster of the star-forming dwarf galaxy NGC 5253. *Astrophys. J.,* **457**, 610-615 (1996).

22. Krumholz, M. R. & McKee, C. F. A general theory of turbulence-regulated star formation, from spirals to ultraluminous infrared galaxies. *Astrophys. J.,* **630**, 250-268 (2005).

23. Caldwell, N., & Phillips, M. M. Star formation in NGC 5253. *Astrophys. J.,* **338**, 789-803 (1989).

24. Kobulnicky, H. A., & Skillman, E. D. Inflows and Outflows in the Dwarf Starburst Galaxy NGC 5253: High-Resolution HI Observations. *Astron. J,*, **135**, 527-537 (2008).

25. López-Sánchez, Á. R., et al. The Intriguing HI gas in NGC 5253: an infall of a diffuse, low-metallicity HI cloud? *Mon. Not. R. Astron. Soc.*, **419**, 1051-1069 (2012).

26. Cen, R. Evolution of cold streams and the emergence of the Hubble sequence. *Astrophys. J.*, **789**, L21 (2014).

27. Harris, J., Calzetti, D., Gallagher, J. S. III, Smith, D. A., & Conselice, C. J. The recent cluster formation histories of NGC 5253 and NGC 3077: Environmental impact on star formation. *Astrophys. J.,* **603**, 503-522 (2004).

28. Cresci, G., Vanzi, L., & Sauvage, M. The star cluster population of NGC 5253. *Astron. & Astrophys*., **433**, 447-454 (2005).

29. de Grijs, R., Anders, P., Zackrisson, E., & Göran, Ö. The NGC 5253 star cluster system – I. Standard modelling and infrared-excess. *Mon. Not. R. Astron. Soc.*, **431**, 2917-2932 (2013).

30. Dekel, A. et al. Cold streams in early massive hot haloes as the main mode of galaxy formation. *Nature*, **457**, 451-454 (2009).

**Acknowledgements** The authors thank John Carpenter, Simon Goodwin, Mark Heyer, Leslie Hunt, Robert Hurt, Mike Jura, Charlie Lada, Claus Leitherer, and Schuyler Van Dyk for assistance with the analysis. The Submillimeter Array is a joint project between the Smithsonian Astrophysical Observatory and the Academia Sinica Institute of Astronomy and Astrophysics and is funded by the Smithsonian Institution and the Academica Sinica.

**Author Contributions** J.L.T., S.C.B., D.J.B., A.K., and D.S.M. performed the observations. J.L.T., S.C.B., and P.T.P.H conceived the project and wrote the observing proposal. J.-H.Z. reduced and imaged the SMA 870μm data; A.K. reduced, imaged, and analyzed the SHARC 350 μm data. J.L.T. and S.M.C. obtained derived quantities and performed data analysis. J.L.T. wrote the first draft, and constructed figures. All authors read, discussed and commented on the draft.

**Author Information** Reprints and permissions information is available at www.nature.com/reprints. Correspondence and requests for materials should be addressed to J.L.T. (e-mail: turner@astro.ucla.edu). The authors declare no competing financial interests.

## Methods

**SMA observations.** NGC 5253 was observed with the Submillimeter Array (SMA)[31] on 17 April 2011. The observing frequency was $\nu_{LO}$=340.323 GHz with 48 adjacent spectral windows covering 4 GHz bandwidth for each of two sidebands. The CO J=3-2 rotational transition at $\nu_0$=345.79599 GHz was in the upper sideband. The array was in the subcompact configuration covering the visibility baselines between 9 and 80 k$\lambda$ corresponding to the angular scales between 29" and 2". Phase center was $\alpha_{J2000}$=13$^h$ 39$^m$ 56.249$^s$, $\delta_{J2000}$ = -31° 38' 29.00". Calibration and reduction was with MIRIAD[32]. The instrumental bandpass was corrected using the quasar 3C 279; complex gains were calibrated using the nearby quasar J1316-336; the flux density scale was determined from the planet model of Neptune. Continuum and line emission were separated using the task *UVLIN* by fitting a linear model to line-free channels. The result is a cube of 25 channels of 10 km/s and a continuum map with effective bandwidth of 8 GHz, convolved to a beam 4" x 2", p.a.=0°, shown in Extended Data Figure 1. Final noise levels were 3 mJy/beam in the continuum map, and 50 mJy/beam in the individual 10 km/s channels.

**CSO SHARC observations.** The 350 μm continuum observations were made with the SHARC camera[33] at the Caltech Submillimeter Observatory on 11–12 January 1999, with 225 GHz opacities around 0.035 and 0.075 for the respective dates. The data consist of 2.2 hours of on-the-fly mapping with a ~60" chopping secondary at 4.132 Hz, and were reduced using CRUSH[34], using an enhanced implementation of the Emerson II deconvolution algorithm[35] which uses sky rotation to fill in the poorly-sampled spatial frequencies of the dual-beam chop. CRUSH removes DC detector offset and correlated sky-noise residuals, flatfields detectors based on sky response, and performs noise weighting, whitening, and despiking. The main beam is 9" FWHM at 350 μm, but the image presented here was smoothed to 12.7" resolution. From observations of Mars taken immediately before the SHARC observations, at a similar elevation, it is estimated that the pointing is good to ~5" rms. The systematic aperture flux calibration of the 350μm image is estimated to be good to within 7% rms.

**Relation of NGC 5253 to M83 and distance.** NGC 5253 is a dwarf spheroidal galaxy of the Cen A/M83 galaxy complex[36] with stellar mass of ~1.5 x 10$^8$ M$_\odot$[37] and estimated[38] total mass, including dark matter, about ten times higher. It is close to the large spiral galaxy M83 in projection. However the distance to M83, 4.8 Mpc[39], is significantly larger than the distance to NGC 5253, at 3.8 Mpc[40]. The HI streamer system [24,25] in the halo of NGC 5253 from which the CO streamer appears to emanate strongly suggests that this dwarf galaxy has had some encounter in its past, but M83 does not appear to be responsible.

**Cloud D CO Emission.** The J = 3 level of CO corresponds to an energy $E_u/k$ of 33 K, a temperature that begins to distinguish actively star-forming clumps from giant molecular clouds. Cloud D is bright in CO(3–2), but only weakly detected in CO(2–1) [3], and not at all in CO(1-0) [5]. The total flux of CO(3-2) emission in the galaxy and streamer is 110 ± 20 Jy km/s, about 30% less than the single dish flux [41]; a typical value for local galaxies, since the array configuration is insensitive to structures < 30" in extent; this value is consistent with the extended streamer emission and with the JCMT-SCUBA continuum image[42]. The CO(2–1) image is shown in Extended Data Figure 2, overlaid on the SMA CO(3–2) image.

CO(3-2) was not detected in previous SMA observations [11] due to insufficient signal-to-noise. Located at $\alpha_{J2000} = 13^h 39^m 55.943^s \pm 0.003^s$, $\delta_{J2000} = -31° 38' 25.097" \pm 0.05"$, the Cloud D CO(3-2) source is coincident to within ~ 0.5" with the core of the supernebula as defined by high-brightness 7mm free-free emission[43]. The CO(3-2) line center is at heliocentric velocity of 397.5±0.6 km s$^{-1}$ and the CO flux of Cloud D is 41 Jy km/s. The size of the CO source in the integrated intensity map deconvolved from the beam is 2.8" × 1.5" ± 7%, p.a. 12° ± 1°. The slight northward extension is consistent with features seen [4,8,21,43,44] in free-free emission, but it is also in the same direction as the elongation of the beams for northern synthesis arrays for this source.

**Cloud D virial mass.** The width of the CO(3-2) line is $\sigma = 9.2 \pm 0.6$ km/s, based on a least squares fit to a Gaussian line profile. We adopt a value for the radius of half the full width half maximum of the geometric mean of the deconvolved source size, using $M_{vir} = \alpha M_\odot$ (v/km/s)$^2$ (r/pc), where v = 2.35$\sigma$, with coefficients of $\alpha = 190$, for $\rho \sim r^{-1}$, adopted here, and $\alpha = 126$ for $\rho \sim r^{-2}$ and 210 for $\rho \sim r^{-0}$ giving the uncertainty limits in $M_{vir}$[45]. We assume that the cloud is turbulently supported against gravity and dispersion-dominated, as for Galactic giant molecular clouds[46,47], hence inclination effects should not be important. If Cloud D is not bound, or has flows that are super-gravitational, then our estimate for the virial mass is an overestimate.

**Supernebula stellar mass.** The stellar mass is based on STARBURST99[48,49] modeling with the following constraints. 1) The 7 mm flux density of the supernebula is 47±4 mJy for the central 2"[8,43]. From this we obtain a Lyman continuum rate of $N_{Lyc} = 7 \times 10^{52}$ s$^{-1}$ for a nebula at 12,000 K[50,51]. 2) The cluster age must be consistent with Brackett γ equivalent width of 255 Å[10,52] and mid-IR ionic line ratios.[53] 3) The cluster must be old enough to have Wolf-Rayet (WR) stars, which would explain the WR spectral signatures[13,14,50,51,54,55]. 4) The cluster must have mass less than the virial mass of $1.8 \times 10^6 M_\odot$, which is also the maximum mass allowed by the [SIV] and Brackett α linewidths[56,57]. We considered both Geneva high and Geneva v=40% breakup stellar models; the high velocity models allow for older cluster ages. The main parameter to vary in fitting the models is the lower mass cutoff to Initial Mass Function (IMF). We consider cluster models with standard Kroupa IMFs with upper mass cutoffs of 150 $M_\odot$, and top-heavy IMFs. Kroupa IMFs with stellar masses down to 0.1 $M_\odot$ cannot give cluster ages sufficiently old to have both WR stars and the given $N_{Lyc}$, given the upper limit on the cluster mass. The IMF must be top-heavy: a cluster of $M_{stars} = 1.1^{+0.7}_{-0.2} \times 10^6 M_\odot$ requires a lower mass cutoff of >3 $M_\odot$. This cluster mass is consistent with previous estimates[58,59]. The Lyman continuum rate inferred from free-free emission may be less than the true value because of leakage from the HII region. In addition, dust can absorb as much as 50% of the ionizing photons in dense Galactic HII regions[60]; this could also increase the stellar mass. Studies of the extended ionized gas in NGC 5253[61] indicate a total galactic star formation rate of twice what we calculate for the supernebula, but this could be ultraviolet photons from nearby, slightly older clusters[27,28,29,58].

**Cloud D continuum and dust mass.** The strong 870 μm continuum source toward Cloud D consists of equal parts free-free emission from the HII region and dust emission. The continuum source is located at $\alpha_{J2000} = 13^h 39^m 55.948^s \pm 0.005^s$, $\delta_{J2000} = -31° 38' 24.88" \pm 0.11"$ (J2000). The 870 μm peak is 0.5" north and 0.32" west of the 7 mm continuum supernebula core[43], which is within the uncertainties of our 4"x 2" beam. The continuum source agrees in position and size with the CO source. The total 870μ flux density of Cloud D is 72 ± 10 mJy. This flux is consistent with previous observations[11], and constitutes ~40% of the total 870 μm flux of 192 mJy for the galaxy as determined from a JCMT/SCUBA map[42]. We extrapolate a free-free flux density from the 7 mm value of 47 ± 4 mJy[8,43], using the $S_\nu \propto \nu^{-0.1}$ spectrum of optically thin emission, giving $S_{870\mu m}^{ff} = 38 ± 4$ mJy for Cloud D. The dust emission is then $S_{870\mu m}^{dust} = S_{870\mu m} - S_{870\mu m}^{ff} = 34 ± 14$ mJy. This value is just consistent with the upper limit set at 1.3 mm [3]. The SHARC flux is $S_{350\mu m}^{dust} = 3.7 ± 0.5$ Jy, of which ~1.0 ± 0.2 Jy originates in Cloud D.

The dust mass is dependent upon the submillimeter dust opacity and dust temperature. We adopt the opacity of the Large Magellanic Cloud [12], extrapolating from κ(160 μm) = 16 cm$^2$ g$^{-1}$ and β = 1.7 to obtain κ(870) = 0.9 cm$^2$ g$^{-1}$. The 350 μm and 870 μm fluxes for Cloud D are consistent with β = 1.7. For dust temperature we adopt $T_{dust}$ = 45 K based on IRAS Point Source Catalog fluxes [62]; $T_{dust}$ < 57 K based on IRAS 60 μm flux. These values of temperature and opacity give $M_{dust} = 1.5 ± 0.1 \times 10^4 M_\odot (T / 45 K)^{-1}$ for the dust mass of Cloud D, with the uncertainty based on the flux. Previous determinations of the dust mass in NGC 5253 were for the entire galaxy, including the streamer, based on the large aperture JCMT/SCUBA[42] flux density; if scaled to our flux of 34 mJy for Cloud D only, these models[59,63] would give a dust mass of $M_{dust}$ ~2-3 × 10$^4$ M$_\odot$, consistent with a cooler dust temperature.

**Dust yield from the cluster and GTD.** The GTD estimated from our dust mass of 1.5 × 10$^4$ M$_\odot$ and our gas mass derived from virial and stellar masses is ~47 (embedded cluster) or 120 (cluster outside cloud). Either is significantly lower than the value of 650 predicted [64] from scaling the Galactic value of 160 [65] to the metallicity of NGC 5253. We argue that the high dust mass is from in situ enrichment by the cluster. The Brackett γ equivalent width of 255 Å[10,52] indicates a cluster age consistent with the presence of WR activity, so mass loss is expected. If the original progenitor cloud had mass ~2 × 10$^6$ M$_\odot$, there would initially have been ~3000 M$_\odot$ of dust in Cloud D, five times less mass than we observe. From STARBURST99 models with z=0.004, Geneva high mass loss stellar models, for a 1.1 × 10$^6$ M$_\odot$ cluster of age 4.4 Myr with an initial mass function of range 3-150 M$_\odot$ and log N$_{Lyc}$ =52.84, one would expect a yield in elements C, N, O, Mg, Si, and Fe of ~24,000 M$_\odot$, of which an estimated 30-50% will be in the form of dust [65,66]. If the cluster is more massive or more top-heavy, the mass loss can be higher. The Geneva rotating models give lower yields, but still sufficient to explain the dust mass. The optical spectrum of NGC 5253 does not yet reflect this localized enrichment: the metallicity of ~0.25 solar is based on nebular lines [13,14] from the extended nuclear HII region, whereas the embedded star cluster is behind at least 16 magnitudes of extinction. However the small CO linewidth suggests that what has been produced in the cloud has so far largely stayed in the cloud.

**Cloud D excitation modeling.** The CO(3-2) to CO(2-1) line ratio can be used to constrain gas density and temperature. Extended Data Figure 2 demonstrates clear differences between the emission in Cloud D and the streamer. RADEX [6] was used to perform non-LTE modeling of the line ratios. We assumed a Black radiation field and column density of 10$^{16}$ cm$^{-2}$, consistent with the observed brightness. Allowed values of density and kinetic temperature for the adopted value of $I_{32}/I_{21}$ = 2.25 and the 1σ lower limit of 2.0 are shown in Extended Data Figure 3. For the value of 2.25, the lower limit on the gas kinetic temperature is $T_K$~350 K; however, the ratio of 2.1 is within the uncertainties and would give $T_K$ > 200 K. The indication of thermal ratios in the near-infrared $H_2$ line ratios [7] would also support a high temperature for this cloud.

**Mass of Cloud D based on $X_{CO}$.** $X_{CO}$ masses are based on CO(1-0) emission, which has not been detected in Cloud D. Given that the CO emission is optically thin, it appears that the $X_{CO}$ value [3] would underpredict the $H_2$ mass by a factor of ~8.

**Streamer kinematics.** The streamer has been detected previously in CO(1-0) and CO(2-1) [3,5]. The emission is found at heliocentric velocities of 410-430 km/s, which is red-shifted by about 20 km/s with respect to the galaxy and the supernebula Cloud D. High resolution VLA images [24] also detect HI emission coincident with this streamer. The streamer coincides with filamentary emission in nebular lines of oxygen[67] and sulfur[68]; it has been suggested that this is an "ionization cone", possibly even due to an AGN[68]. We suggest that this emission is due to leakage of photons from the starburst, which are ionizing the surface of the infalling streamer.

**Streamer CO and dust properties and GTD.** The CO(3-2) emission originates largely from a single cloud, Cloud "C" [3]. For its line strength of $I_{32}$ = 3.9±0.6 K km/s, the ratio of $I_{32}/I_{21}$ is 1.0 ± 0.3. RADEX models of the ratio are consistent with optically thick and cooler gas, T ~ 20 K, with n ~$10^{3.5}$-$10^4$ cm$^{-3}$ (Extended Data Figure 3). The molecular mass of the streamer based on CO(2-1)[3] is $M_{H2}$ = 2 x $10^6$ $M_\odot$ for a Galactic conversion factor, $X_{CO}$ = 2 x $10^{20}$ cm$^{-2}$ (K km s$^{-1}$)$^{-1}$. The virial mass is 3 x $10^6$ $M_\odot$[3]. The 870µm dust continuum emission follows the CO (Figure 2). The 870 µm flux density of the streamer is 26 ± 8 mJy, which is all dust (Figure 2). Adopting the dust opacity κ(870 µm) = 0.9 cm$^2$ g$^{-1}$ and dust temperature $T_d$ = 20 K, we obtain a dust mass of $M_{dust}$= 2.6 x $10^4$ $M_\odot$. The streamer thus has GTD~120 using the virial mass for the gas mass. If the HI gas[24] is added, the total H+$H_2$ mass becomes 4.3 x $10^6$ $M_\odot$, which gives GTD=170. That the streamer is molecular gas, and even more surprisingly, dense molecular gas, is difficult to understand. Molecular gas favors high pressure environments[69] such as the midplanes of the central regions of spiral disks. Even though the filament appears to be in a low pressure environment, it is not only molecular, but dense. Models of the streamer as an example of a primordial cooling filament, in which the gas collapses toward the center of the dark matter potential, are able to produce the observed inflow rate of gas of ~0.1-0.2 $M_\odot$ yr$^{-1}$, but are unable to reproduce the formation of the observed GMCs [9]. The streamer may be previously enriched gas.

**Cloud F.** Cloud F, located about 5" to the southwest of Cloud D, was not detected in previous CO observations [3,5]. Clouds D and F are the only two detected GMCs within NGC5253 proper. Using the Galactic CO conversion factor and assuming optically thick emission, for the observed flux of 17 ± 6 Jy km/s we obtain a mass of $M_{H2}$ ~ 4 x $10^5$ $M_\odot$ for Cloud F.

**Star formation efficiency and cluster survival.** The canonical value of η = 50% is from virial considerations for the survival of a bound cluster with mass loss on timescales less than the crossing time[70]. It is possible for a cluster to survive with lower efficiency, to ~30%, if the gas is lost slowly and the cluster expands[2,71].

31. Moran, J. M., & Ho, P. T. P. Smithsonian Submillimeter Wavelength Array. *Proc. SPIE*, **2200**, 335-346 (1994).

32. Sault, R.J., Teuben, P.J., & Wright, M.C.H. A retrospective view of Miriad. *ASP Conf. Ser.*, **77**, 433-436 (1995).

33. Wang, N., et al. A submillimeter high angular resolution bolometer array camera for the Caltech submillimeter observatory. *Proc. STT*, **7**, 426-438 (1996).

34. Kovács, A. CRUSH: fast and scalable data reduction for imaging arrays. *Proc. SPIE*, **7420**, 45-59 (2008).

35. Emerson, D. T. Approaches to multi-beam data analysis. *ASP Conf. Ser.*, **75**, 309-317 (1995).

36. Karachentsev, I. D. et al. The Hubble Flow around the Centaurus A/M83 Galaxy Complex. *Astron. J.,* **133**, 504-517 (2007).

37. Martin, C. L. The impact of star formation on the interstellar medium in dwarf galaxies. II. The formation of galactic winds. *Astrophys. J.,* **506**, 222-252 (1998).

38. Persic, M., Salucci, P., & Stel, F. The universal rotation curve of spiral galaxies – I. The dark matter connection. *Mon. Not. R. Astron. Soc.,* **281**, 27-47 (1996).

39. Radburn-Smith, D. J. et al. The GHOSTS Survey. I. Hubble Space Telescope Advanced Camera for Surveys Data. *Astrophys. J.*, **195**, 18 (2011).

40. Sakai, S., Ferrarese, L., Kennicutt, R. C., Jr., & Saha, A. The effect of metallicity on Cepheid-based distances. *Astrophy. J.*, **608**, 42-61 (2004).

41. Meier, D. S., Turner, J. L., Crosthwaite, L. P., & Beck, S. C. Warm molecular gas in dwarf starburst galaxies: CO(3-2) observations. *Astron. J.,* **121**, 740-752 (2001).

42. James, A., Dunne, L., Eales, S., & Edmunds, M.G. SCUBA observations of galaxies with metallicity measurements: a new method for determining the relation between submillimetre luminosity and dust mass. *Mon. Not. R. Astron. Soc.,* **335**, 753-761 (2002).

43. Turner, J. L., & Beck, S. C. The birth of a super star cluster. *Astrophys. J.*, **602**, L85-L88 (2004).

44. Turner, J. L., Ho, P. T. P., & Beck, S. C. The radio properties of NGC 5253 and its unusual HII regions. *Astron. J.,* **116**, 1212-1220 (1998).

45. MacLauren, I., Richardson, K. M., & Wolfendale, A. W. Corrections to virial estimates of molecular cloud masses. *Astrophys. J.*, **333**, 821-825 (1988).

46. Solomon, P. M., Rivolo, A. R., Barrett, J., & Yahil, A. Mass, luminosity, and line width relations of Galactic molecular clouds. *Astrophys. J.*, **319**, 730-741 (1987).

47. Heyer, M., & Brunt, C. M. The universality of turbulence in Galactic molecular clouds. *Astrophys. J.*, **615**, L45-L48 (2004).

48. Leitherer, C., et al. Starburst99: Synthesis Models for Galaxies with Active Star Formation, *Astrophys. J. Suppl.,* **123**, 3-40 (1999).

49. Leitherer, C., et al. The effects of stellar rotation. II. A comprehensive set of STARBURST99 models. *Astrophys. J. Suppl.*, **212**, 18 (2014).

50. Walsh, J. R., & Roy, J.-R. Optical spectroscopic and abundance mapping of the amorphous galaxy NGC 5253, *Mon. Not. R. Astron. Soc.,* **239**, 297-324 (1989).

51. Monreal-Ibero, A., Walsh, J. R., & Vílchez, J. M. The ionized gas in the central region of NGC 5253. 2D mapping of the physical and chemical properties. *Astron. Astrophys.* **544**, A60 (2012).

52. Davies, R. I., Sugai, H., & Ward, M. J. Star-forming regions in blue compact dwarf galaxies. *Mon. Not. R. Astron. Soc.,* **295**, 43-54 (1998).

53. Martín-Hernandez, N. L., Schaerer, D., Sauvage, M. High spatial resolution mid-infrared spectroscopy of NGC 5253: The stellar content of the embedded super-star cluster", *Astron. Astrophys.,* **429**, 449-467 (2005).

54. Schaerer, D., Contini, T., Kunth, D., & Meynet, G. Detection of Wolf-Rayet stars of WN and WC subtypes in super-star clusters of NGC 5253. *Astrophys. J.,* **481**, L75-L79 (1997).

55. Westmoquette, M. S., James, B., Monreal-Ibero, A., Walsh, J. R. Piecing together the puzzle of NGC 5253: abundances, kinematics, and WR stars. *Astron. & Astrophys.,* **550**, A88 (2013)

56. Turner, J. L., et al. An extragalactic supernebula confined by gravity. *Nature*, **423**, 621-623 (2003).

57. Beck, S. C., et al. [SIV] in the NGC 5253 supernebula: ionized gas kinematics at high resolution. *Astrophys. J.,* **755**, 59 (2012).

58. Calzetti, D. et al. Dust and recent star formation in the core of NGC 5253. *Astron. J.,* **114**, 1834-1849 (1997).

59. Vanzi, L., & Sauvage, M. Dust and super star clusters in NGC 5253. *Astron. & Astrophys*. **415**, 509-520 (2004).

60. McKee, C. F., & Williams, J. P. The luminosity function of OB associations in the Galaxy. *Astrophys. J.*, **476**, 144-165 (1997).

61. Martin, C. L. & Kennicutt, R. C., Jr. Soft X-ray emission from NGC 5253 and the ionized interstellar medium. *Astrophys. J.*, **447**, 171-183 (1995).

62. Thronson, H. A., & Telesco, C. M. Star formation in active dwarf galaxies. *Astrophys. J.*, **311**, 98-112 (1986).

63. Hunt, L., Bianchi, S., & Maiolino, R. The optical-to-radio spectral energy distributions of low-metallicity blue compact dwarf galaxies. *Astron. & Astrophys.*, **434**, 849-866 (2005).

64. Rémy-Ruyer, A. et al. Gas-to-dust ratios in local galaxies over a 2 dex metallicity range. *Astron. Astrophys.*, **562**, A31 (2014).

65. Zubko, V., Dwek, E., & Arendt, R. G. Interstellar dust models consistent with extinction, emission, and abundance constraints. *Astrophys. J. Suppl.,* **152**, 211-249 (2004).

66. Jenkins, E. B. A unified representation of gas-phase element depletions in the interstellar medium. *Astrophys. J.,* **700**, 1299-1348 (2009).

67. Graham, J. Filamentary Structure in NGC 5253. *Publ. Astron. Soc. Pac.,* **93**, 552-553 (1981).

68. Zastrow, J., Oey, M. S., Veilleux, S., McDonald, M., & Martin, C. L. An ionization cone in the dwarf starburst galaxy NGC 5253. *Astrophys. J.*, **741**, L17 (2011).

69. Elmegreen, B. A pressure and metallicity dependence for molecular cloud correlations and the calibration of mass. *Astrophys. J.*, **338,** 178-196 (1989).

70. Mathieu, R. D. Dynamical constraints on star formation efficiency. *Astrophys. J.,* **267**, L97-L101 (1983).

71. Baumgardt, H., Kroupa, P., & Parmentier, G. The influence of residual gas expulsion on the evolution of the Galactic globular cluster system and the origin of the Population II halo. *Mon. Not. R. Astron. Soc.,* **384**, 1231-1241 (2008).

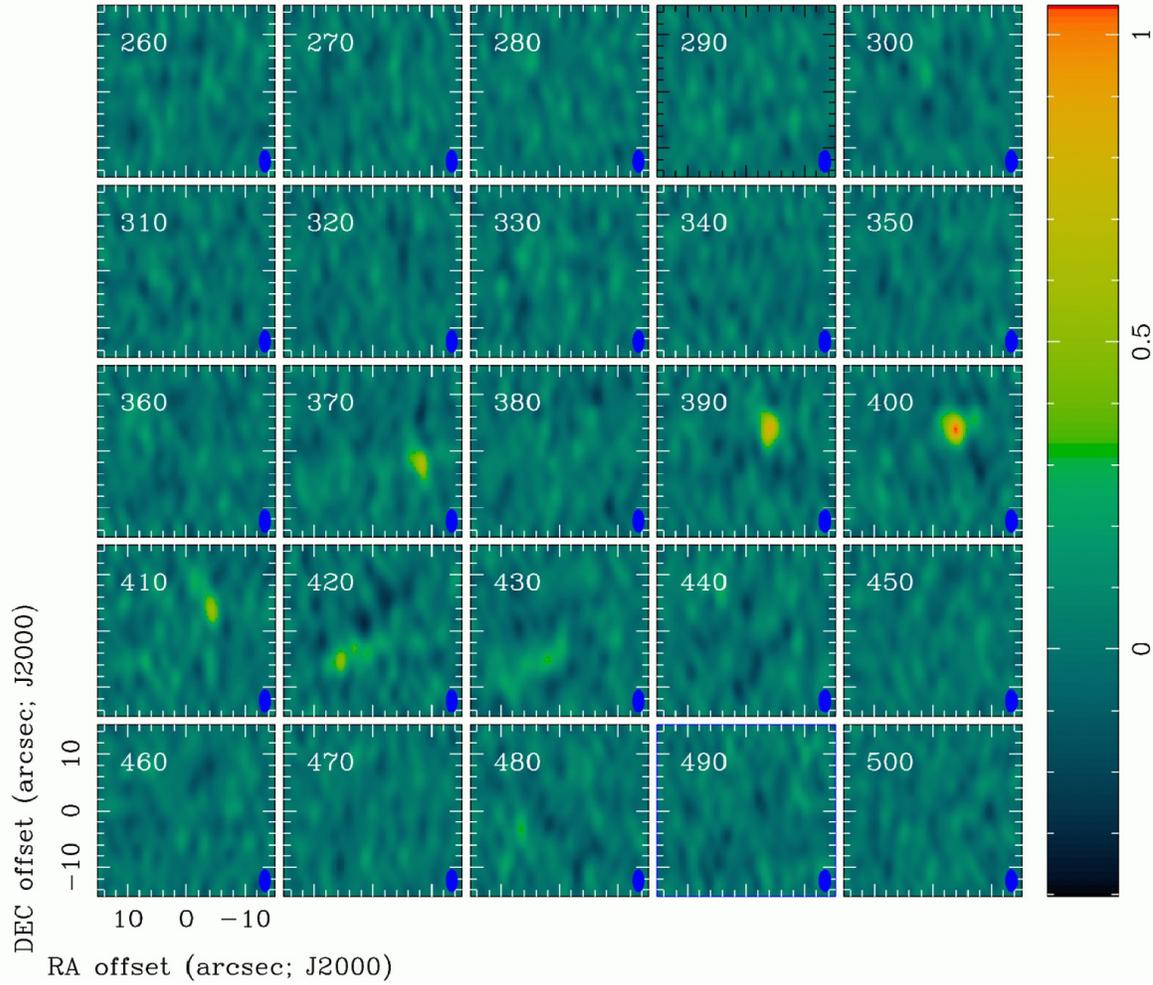

**Extended Data Figure 1. Channel maps of CO(3-2) emission in NGC 5253.** Positions are relative to a reference position 13$^h$ 39$^m$ 56.249$^s$, -31° 38' 29" (J2000). Channels are 10 km/s wide; the heliocentric velocity is noted on the individual maps. The color bar range maximum is 1 Jy/beam for each 10 km/s channel. The beam is 4" x 2", p.a. 0°.

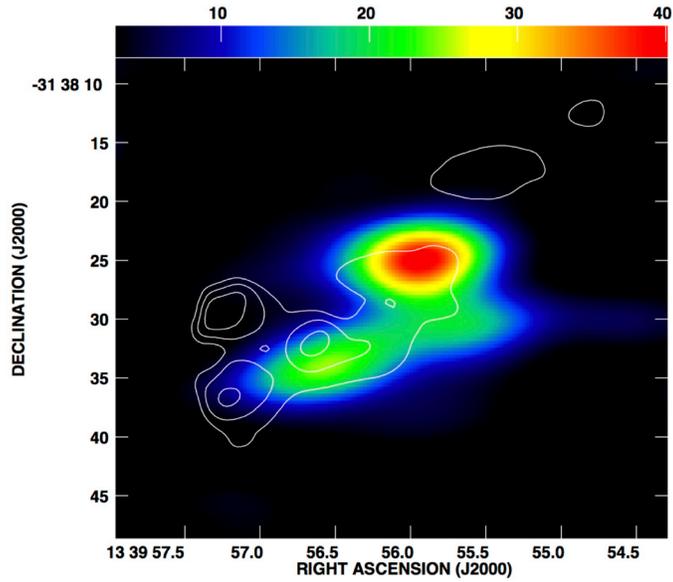

**Extended Data Figure 2. SMA image of CO(3-2) with CO(2-1).** CO(3-2) emission is shown in color, with an image[2] from the Owens Valley Millimeter Array of CO(2-1) in contours. The SMA CO(3-2) image has been smoothed from its original 4" x 2", p.a. 0° resolution to match the 9.7" x 5", p.a. -84° beam of the CO(2-1) image. Color image flux range is 3 to 40 Jy/bm km/s. Contours are linear multiples of 4 Jy/bm km/s.

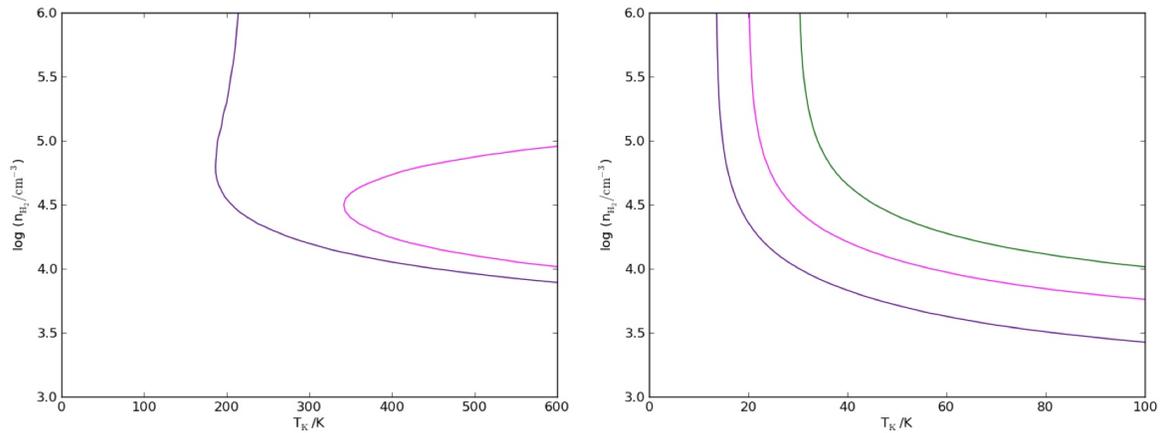

**Extended Data Figure 3. RADEX modeling of Cloud D and Streamer.** Escape probability transfer modeling of the CO(3-2) to CO(2-1) line ratio. Models were run using a Black radiation field, spherical escape probability and $N_{co} = 10^{16}$ cm$^{-2}$. (a) Cloud D. Magenta is for the value $I_{32}/I_{21} = 2.25$, the optically thin limit, and indigo is for the 1σ lower limit to the measured value of 2.6. There is no solution for $I_{32}/I_{21} = 2.6$. (b) Streamer. Magenta is for $I_{32}/I_{21} = 0.98$, and indigo and green are solutions for the ± 1σ values respectively.